\documentclass[%
reprint, pdftex
amsmath,
amssymb,
aps,
prx
showpacs
]{revtex4-1}

\usepackage{graphicx}% Include figure files
\usepackage{dcolumn}% Align table columns on decimal point
\usepackage{bm}% bold math
\usepackage[usenames,dvipsnames,svgnames,table]{xcolor}
\usepackage[normalem]{ulem}
\usepackage{upgreek}

\begin{document}
	
	\preprint{APS/123-QED}
	\title{Lasing on a narrow transition in a cold thermal strontium ensemble}% 
	\author{Stefan A. Sch\"{a}ffer, Mikkel Tang, Martin R. Henriksen, Asbj\o rn A. J\o rgensen, Bjarke T. R. Christensen, and Jan W. Thomsen}%
	\affiliation{Niels Bohr Institute, University of Copenhagen, Blegdamsvej 17, 2100 Copenhagen, Denmark}

	\begin{abstract}
		Highly stable laser sources based on narrow atomic transitions provide a promising  platform for direct generation of stable and accurate optical frequencies. Here we investigate a simple system operating in the high-temperature regime of cold atoms. The interaction between a thermal ensemble of $^{88}$Sr at mK temperatures and a medium-finesse cavity produces strong collective coupling and facilitates high atomic coherence which causes lasing on the dipole forbidden $^1$S$_0 \leftrightarrow ^3$P$_1$ transition. We experimentally and theoretically characterize the lasing threshold and evolution of such a system, and investigate decoherence effects in an unconfined ensemble. We model the system using a Tavis-Cummings model, and characterize velocity-dependent dynamics of the atoms as well as the dependency on the cavity-detuning. %These investigations allow identification of important parameters for a future active optical clocks.
	\end{abstract}
	
	% PACS, the Physics and Astronomy
	% Classification Scheme.
	%Use showkeys class option if keyword
	%display desired
	\maketitle
	
	%\tableofcontents
	\section{Introduction}
	Active optical clocks have been suggested as an excellent way to improve the short-time performance of optical clocks by removing the requirement for an ultra stable interrogation laser \cite{Chen2005, Chen2009, MeiserPRL2009}. Current state-of-the-art neutral-atom optical lattice clocks with exceedingly high accuracy are typically limited in precision by the interrogation lasers and not by the atomic quantum projection noise limit. The high Q-value of the atomic transitions ($Q>10^{17}$) exceeds that of the corresponding reference laser stability for the duration of the interrogation cycle \cite{Oelker, McGrew, Nicholson, Ushijima, Huntemann}. State-of-the-art reference lasers now perform below the level of $10^{-16}$ fractional frequency stability at $1$~s \cite{Matei, Zhang, Robinson}. This recent advance in performance is enabled by using crystalline mirror coatings and spacers as well as employing cryogenic techniques. Nevertheless these lasers continue to be limited by the thermal noise induced in the reference cavity mirrors \cite{Numata, Cole}. By using an active optical clock, the narrow spectral features of the atoms themselves produce the reference light needed, via direct generation of lasing in an optical cavity \cite{Laske}. Operation in the bad-cavity limit, where the cavity field decays much faster than the bare atomic transition, allows high suppression of the cavity noise in these lasers \cite{NorciaPRX2018}. It has been predicted that such systems could reach Q-factors in excess of the transition Q-factor due to narrowing caused by collective effects \cite{MeiserPRL2009,Debnath}. It was recently shown experimentally that an active optical atomic clock can perform at the $10^{-16}$ fractional frequency stability level between $1-100$~s, and with a fractional frequency accuracy of $4\times10^{-15}$ by using a cyclically operated optical lattice with $^{87}$Sr atoms \cite{NorciaPRX2018}. This is a performance level in stability also promised \cite{Christensen} or demonstrated \cite{Olson} by other types of thermal atom systems. Continuous lasing with atoms at room temperature have shown performance at the $10^{-13}$ level of fractional frequency stability \cite{Shi}.
	
	These results demonstrate the potential of active atomic clocks, and motivates the present attempt to improve the understanding of atom dynamics in such a system. We concentrate on a cold atomic gas consisting of bosonic $^{88}$Sr cooled to a temperature of $T\approx5$~mK, and permitted to expand freely as a thermal gas while lasing. Superradiant lasing in such a system would substantially reduce the technological challenges of maintaining truly continuous operation compared to the case of atoms confined in an optical lattice trap \cite{NorciaPRX2016, Kazakov}.
	
	We experimentally demonstrate pulsed lasing on a narrow transition in $^{88}$Sr and describe the characteristic dynamics of our system. Using a Tavis-Cummings model we simulate the full system consisting of up to $8\times10^7$ individual atoms and give a qualitative explanation of the dynamics. Rich velocity- and position-dependent dynamics of the atoms are included in the description and become an important factor for understanding the lasing behavior. The model allows us to quantify these behaviors and set requirements on ensemble size and temperature to realize strongly driven or weakly driven superradiance respectively. Such requirements are useful for the future development of truly continuous lasing on ultra-narrow clock transitions in strontium, or other atomic species.

	In section II we describe the physical characteristics of our system and the experimental routine. Section III describes our theoretical model for the cold-atom based laser, and the proposed lasing dynamics. In Section IV the lasing characteristics are presented and comparisons between experiment and simulations are made.
	
	\section{Experimental system}
	
\begin{figure}[t]
		\includegraphics[width=\columnwidth]{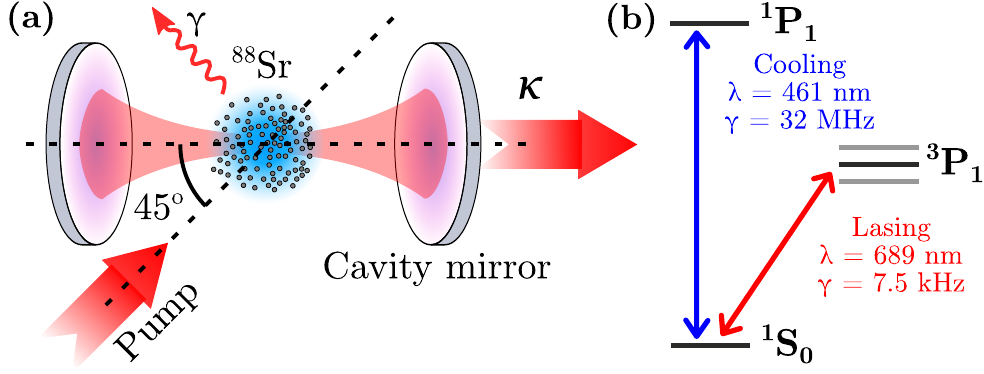}
		\caption{\textbf{(a)} Schematic of the experimental system showing a thermal ensemble of strontium atoms partially overlapping with the cavity mode. Pumping prepares the atoms in the excited state, and allows subsequent emission of a coherent pulse into the cavity mode. Light couples out of both ends of the cavity with the cavity decay rate $\kappa$, and is detected at one end only. \textbf{(b)} Level scheme for $^{88}$Sr showing the cooling and lasing transitions. Wavelength ($\lambda$) and natural transition decay rates ($\gamma$) are indicated.\label{fig:setup}}
	\end{figure}

	The system consists of an ensemble of cold $^{88}$Sr atoms cooled and trapped in a 3D Magneto-Optical Trap (MOT) from a Zeeman-slowed atomic beam, using the $^1$S$_0 \leftrightarrow ^1$P$_1$ transition, and was described in \cite{Schaffer}. The atomic cloud partially overlaps with the mode of an optical cavity, and can be state manipulated by an off-axis pumping laser, see Fig. \ref{fig:setup}. The cavity can be tuned on resonance with the $^1$S$_0 \leftrightarrow ^3$P$_1$ intercombination line, and has a linewidth of $\kappa = 2\pi\cdot 620$~kHz. Our mirror configuration ensures a large cavity waist radius of $w_0 = 450$~$\upmu$m, which ensures a reasonably high intra-cavity atom number, $N_{cav}$ of order $1\times10^7$. $N_{cav}$ is estimated from fluorescence measurements and lasing pulse intensities, and given by $N_{cav}\equiv \eta_{cav} N$, where $\eta_{cav}\simeq 0.15$ is a geometric factor corresponding to the fraction of atoms within a cylinder of radius $w_0$ along the cavity mode. Since all atoms have different velocities and positions, the cavity coupling $g_c^j$ for the $j$'th atom is not constant. While this is included in the theoretical simulations, for discussion of regimes and general scaling behavior we use an effective coupling to the cavity mode averaged over the time-dependent position of the atoms $g_\textit{\scriptsize{eff}}\equiv\langle |g_i|\rangle=g_c^0\frac{1}{N}\sum_{i=1}^{N}|\zeta_z(z_i)|\zeta_{xy}(x_i,y_i)$.
	
	The large cavity waist ensures a negligible decoherence induced from transit time broadening of $\Gamma_{tt} = 2\pi\cdot 2.2$~kHz. While the $^1$S$_0 \leftrightarrow ^3$P$_1$ transition used for operation has a linewidth of $\gamma =2\pi\cdot 7.5$~kHz, the ensemble temperature $T$ causes a Gaussian velocity distribution of width $\sigma_v= 0.69$~m/s and a Doppler broadened FWHM of $\Gamma_D = 2\pi\cdot2.3$~MHz. The bare atoms are thus deep in the bad-cavity regime, $\gamma\ll\kappa$, whereas the total inhomogeneously broadened ensemble is just below the bad-cavity threshold. Here, the cavity field decay rate and total atomic decoherence rate is of similar size $\kappa\sim\Gamma_{dec}$, and ensemble preparation can heavily affect the lasing process and the attainable suppression of cavity noise. We characterize the system by its collective cooperativity $C_N=C_0 N$, and the single-atom cooperativity is given by $C_0=\frac{(2g_\textit{\scriptsize{eff}})^2}{\kappa\gamma}$. This leads us to the definition of a collective coupling rate $\Omega_N=2\sqrt{N}g_\textit{\scriptsize{eff}}$. While the spectral linewidth of the emitted light is controlled by $C_0$ \cite{MeiserPRL2009,Kuppens}, the coherence build-up, and thus lasing power, is determined by $C_N$. Increasing the coupling rate $g_\textit{\scriptsize{eff}}$, and thus the single-atom cooperativity, will then improve interaction at the expense of the ultimately attainable lasing linewidth.
	
	After preparing the MOT, the cooling light is switched off, and an off-axis pumping beam is used to excite the atoms to the $^3$P$_1$ state. By pumping the atoms on resonance for $170$~ns a peak excitation of approximately $85~\%$ is obtained for the atoms within the cavity mode. Inhomogeneous Doppler detuning caused by the thermal distribution of the atoms, and the large spatial distribution of the full atomic ensemble with respect to the pumping field result in varying Rabi frequencies for different atoms. The collective atomic Bloch vector will thus contain some level of atomic coherence from the imperfect pump pulse. Due to the excitation angle of $45^\circ$ the pump phase periodicity along the cavity axis suppresses any forced coherence between the atoms and cavity field. Additionally, the spatial coherence of the fast atoms will wash out as they propagate inhomogeneously along the cavity axis. In order to remove any remaining phase-coherence we apply light at $461$~nm detuned about $\Delta\nu=-41~$MHz from the $^1$S$_0 \leftrightarrow ^1$P$_1$ for 500 ns after pumping. This results in an average of $1.3$ scattering events per atom, and ensures atoms are either in the excited or ground state. We see no quantifiable change in the lasing behavior by doing this.
	
	Once excited, the atoms will build up coherence mediated through the cavity field and emit a coherent pulse of light into the cavity mode. Light leaking through one of the cavity mirrors is then detected on a photodiode. We couple a reference field, $s$, into a cavity mode far off resonance, $\Delta_{FSR}=781$~MHz, with the $^1$S$_0 \leftrightarrow ^3$P$_1$ transition, allowing us to lock the cavity length on resonance with the atoms.
	
	\begin{figure}[t]
		\includegraphics[width=\columnwidth]{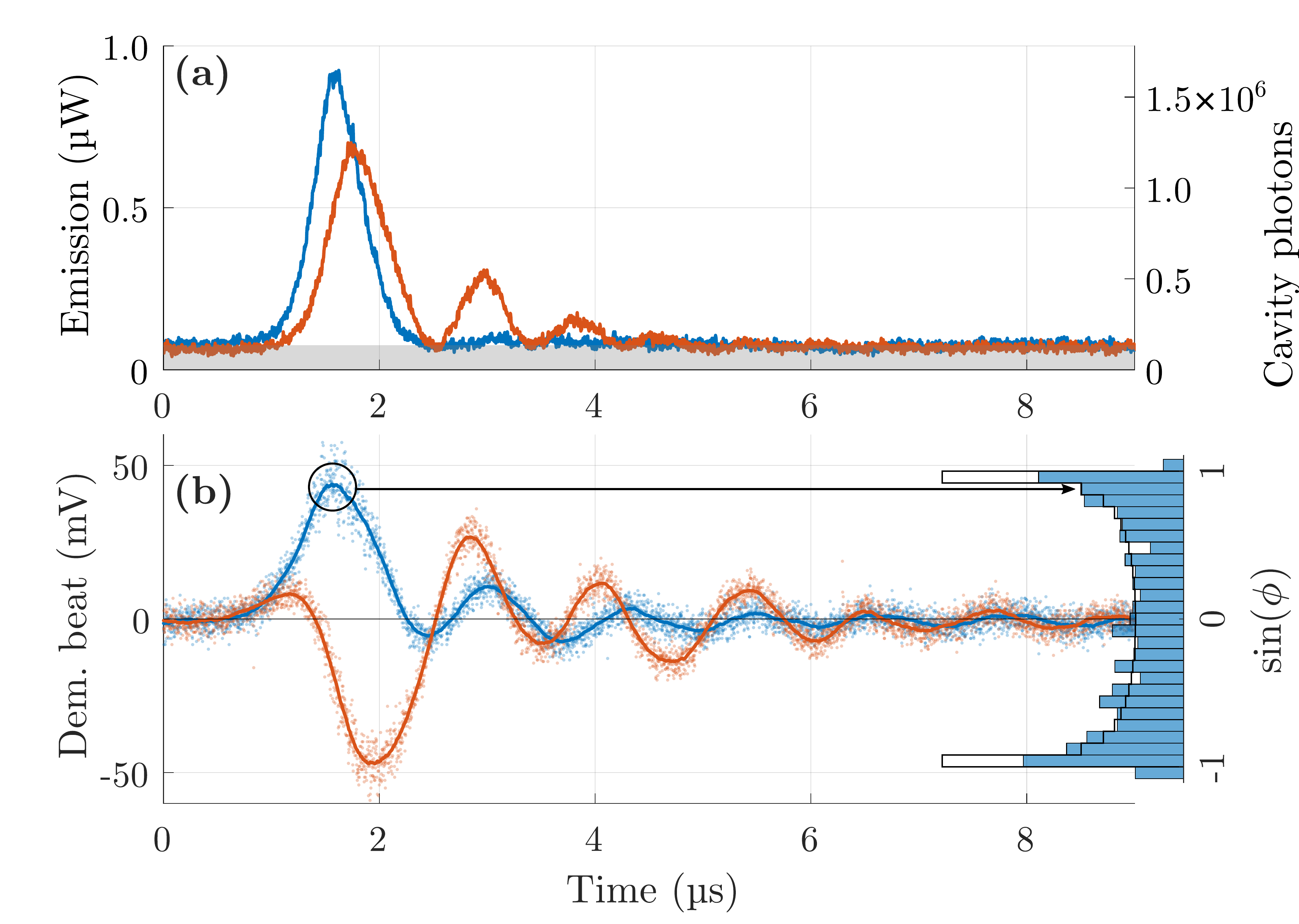}
		\caption{Experimental data showing the time evolution of a lasing pulse after pumping for cavity-atom detuning $\Delta_{ce}=0$~kHz (blue) and $\Delta_{ce}=900$~kHz (orange) respectively. The pumping pulse ends at time $t=0~\upmu$s, and the atom number used is $N=7.5\cdot10^7$. \textbf{(a)} A background power level of $75$~nW in shaded gray indicates the constant non-interacting reference laser field. \textbf{(b)} Beat signal between reference field and laser pulses. The dots are data from two individual samples whose running mean is shown with full lines. The right-hand inset shows a histogram over 740 datasets of the distribution of phase-values at maximum pulse intensity for the case of $\Delta_{ce}=0$~kHz. An ideal homogeneous phase-distribution would give the histogram outlined in black. \label{fig:Pulse}}
	\end{figure}
	
	Fig. \ref{fig:Pulse} shows an example of a lasing pulse and the associated beat signal when the cavity is on resonance (blue) and detuned $\Delta_{ce}=900$~kHz (orange) from the atomic resonance respectively. The pumping pulse ends at $t=0$~s and after some delay, $\tau$ (on the order of few $\upmu$s), a lasing pulse is emitted into the cavity mode. The lasing pulses are followed by a number of revivals of the superradiant output. This is a fingerprint of the coherent evolution of the cavity-atom system, where light emitted into the cavity mode is reabsorbed by the atoms, and is then re-emitted into the cavity mode. The lasing pulse shown in Fig. \ref{fig:Pulse} (a) is detected superimposed on a $75$~nW background. This background is a reference field used for locking the length of the cavity, and provides the opportunity for a heterodyne beat with the lasing pulse. At zero cavity-atom detuning ($\Delta_{ce}=0$) the oscillating revivals are barely visible due to reference field intensity noise. When detecting the beat signal, however, the signal scales linearly rather than quadratically with the lasing E-field $E_{las}$, and here multiple revivals are visible in both the resonant and off-resonant cases. 
	
	The beat signal is given by $S_{beat}\propto|E_{las}|\sin{\phi}$. The envelope thus gives us an improved signal-to-noise ratio at low field values as compared with the emitted intensity. We down-sample the signal by beating it with the frequency $\Delta_{FSR}$, and $\sin{\phi}$ then allows us to retrieve the phase information. In Fig. \ref{fig:Pulse} (b) we see the phase during the initial lasing pulse differ between the two datasets. Because the lasing frequency does not follow the cavity detuning $\Delta_{ce}$ in a bad cavity our demodulation frequency is slightly off, and the phase becomes $\Phi=\phi+(1-\xi_\textit{\scriptsize{pull}})\Delta_{ce}t$ where $\xi_\textit{\scriptsize{pull}}$ is the cavity frequency pulling factor. This results in the additional zero-crossings in the beat signal of the $\Delta_{ce}=900$~kHz signal as compared to the emission intensity, e.g. during the primary pulse. Additionally, we observe that the phase $\phi$ is random between experimental cycles at a given detuning as a result of the randomized phase of the atomic coherence for each realization of the pulse, see inset in Fig. \ref{fig:Pulse} (b).
		
	\section{Semi-classical model}
	In order to gain an improved understanding of the dynamics involved in our system, we simulate the behavior with a Monte-Carlo approach. We model the system using a Tavis-Cummings Hamiltonian for an N-atom system. The description includes a classical pumping field at frequency $\omega_p$. In the Schr\"{o}dinger picture the full expression becomes:
	\begin{eqnarray}
		H &=& \hbar \omega_c a^\dagger a + \sum_{j=1}^N  \hbar \omega_{e} \sigma_{ee}^j \\\nonumber
		%&+& \frac 1 2 \hbar \eta \left( a e^{ -i \omega_s t } + a^{\dagger} e^{ i \omega_s t } \right)\\\nonumber
		&+& \sum_{j=1}^N \hbar g_c^j  \left( \sigma_{ge}^j + \sigma_{eg}^j \right) \left( a + a^\dagger \right)\\\nonumber
		&+& \sum_{j=1}^N \hbar \frac{\chi_p^j}{2}  \left( \sigma_{ge}^j + \sigma_{eg}^j \right) \left( e^{ i \vec{k}_p \cdot \vec{r}_j - i \omega_p t } + e^{ -i \vec{k}_p \cdot \vec{r}_j + i \omega_p t } \right).
	\end{eqnarray}
	
	Here $\omega_c$ is the angular frequency of the cavity mode, $a$ is the corresponding lowering operator. The involved electronic energy states $|g\rangle$ and $|e\rangle$ correspond to the ground and excited atomic states, with a transition frequency of $\omega_e$, as seen in Fig. \ref{fig:setup} (b). The time-dependent position of the atom is given by $\vec{r}_j=(x_j, y_j, z_j)$, and $\vec{k}_p$ is the pump beam wavevector. The pump beam has a semi-classical interaction term with Rabi frequency $\chi_p^j$ and the interaction between cavity field and the j'th atom is governed by the coupling factor $g_c^j$:
	\begin{eqnarray}
	g_c^j &=& g_c^0\cdot\zeta_z(z_j)\cdot\zeta_{xy}(x_j,y_j)\nonumber\\
	&=& \sqrt{\frac{6c^3\gamma}{w_0^2L\omega_{e}^2}}\cdot\sin(k_e z_j)\cdot e^\frac{x_j^2+y_j^2}{w_0},
	\end{eqnarray}
	where L is the cavity length and the wave number $k_e=\omega_{e}/c$. The Doppler effect is included in the model through the time-dependencies of the atom positions and, in particular for interaction with the cavity mode, through the resulting variation in $g_c^j(t)$ caused by $\zeta_z(z_j(t))=\sin(k_e z_j(t))$.
	%and $\eta=\sqrt{P_{in}\kappa/\hbar\omega_c}$ is the rescaled power of the classical field.
	
	By entering an interaction picture, and using the rotating wave approximation, the time evolution of the system operators can be obtained. Here we make the semi-classical approximation of factorizing the expectation values for operator products, which results in linear scaling of the number of differential equations with the number of atoms. This approximation results in the neglect of all quantum noise in the system, and by consequence any emerging entanglement \cite{Gonzales-Tudela, Wolfe,Barberena}. We motivate this assumption by the very large number of atoms in the system, whose individual behaviors are taken into account with separate coupling factors $g_c^j$. The quantum noise is then expected to be negligible compared to the single-operator mean values. The operator mean values are described by three distinct forms of evolution:
	
	\begin{eqnarray} \label{eq:OBE}
		\dot{\left\langle \sigma_{ge}^j \right\rangle} &=& -\left( i\Delta_{ep} + \frac{\Gamma}{2} \right) \left\langle \sigma_{ge}^j \right\rangle \\\nonumber
		&&+ i \left( g_c^j \langle a \rangle + \frac { \chi_p^j}{2} e ^ { -i \vec{k}_p \cdot \vec{r}_j } \right) \left( \left\langle \sigma_{ee}^j \right\rangle - \left\langle \sigma_{gg}^j \right\rangle \right)\\\nonumber
		\dot{\left\langle \sigma_{ee}^j \right\rangle} &=&
		- \Gamma \left\langle \sigma_{ee}^j \right\rangle + 
		i \left( g_c^j \left\langle a^\dagger \right\rangle + \frac{\chi_p^j}{2} e^{ i \vec{k}_p \cdot \vec{r}_j} \right) \left\langle \sigma_{ge}^j \right\rangle\\\nonumber
		&&- i \left( g_c^j \langle a \rangle + \frac{\chi_p^j}{2} e^{ -i \vec {k}_p \cdot \vec{r}_j } \right) \left\langle \sigma_{eg}^j \right\rangle.\\\nonumber
		\dot {\langle a \rangle} &=& %- \frac i 2 \eta e^{ i \Delta_{sp} t } 
		- \left( i \Delta_{cp} + \frac { \kappa } { 2 } \right) \langle a \rangle 
		- \sum_{j=1}^N i g_c^j \left\langle \sigma_{ge}^j \right\rangle.
	\end{eqnarray}
	Here $\Delta_{n m} = \omega_n - \omega_m$ is the detuning of the n'th field with respect to the m'th field. Because of the semi-classical approximation the hermitian conjugate of these operators are simply described by the complex conjugate of their mean values, while $\langle\sigma_{ee}^j\rangle + \langle\sigma_{gg}^j\rangle =1$. This leaves us with a total of $1+2N$ coupled differential equations where $N$ is on the order of $10^7$.
	
	The Monte-Carlo simulation assumes the atoms are initially in the ground state, and subsequently pumped into the excited state and left to evolve with time. All parameters are estimated experimentally and there are no explicit free fitting parameters (see appendix). Atomic positions and velocities are sampled randomly from a 3D Gaussian and a thermal Maxwell-Boltzmann distribution respectively. The velocities are distributed according to a temperature of $T=5$~mK and the positions with a standard deviation of $0.8$~mm. Atomic motion is treated classically and without collisions. The atoms interact only via the cavity field, and any spontaneous emission into the cavity mode is neglected. The lasing process is initiated by an initial nonzero total coherence $\sum_{j=1}^N \sigma_{ge}^j$, resulting from the pumping pulse. This replaces the role of quantum noise in the system, and without it the system would couple only to the reservoir.
	
\begin{figure}[t]
		\includegraphics[width=0.9\columnwidth]{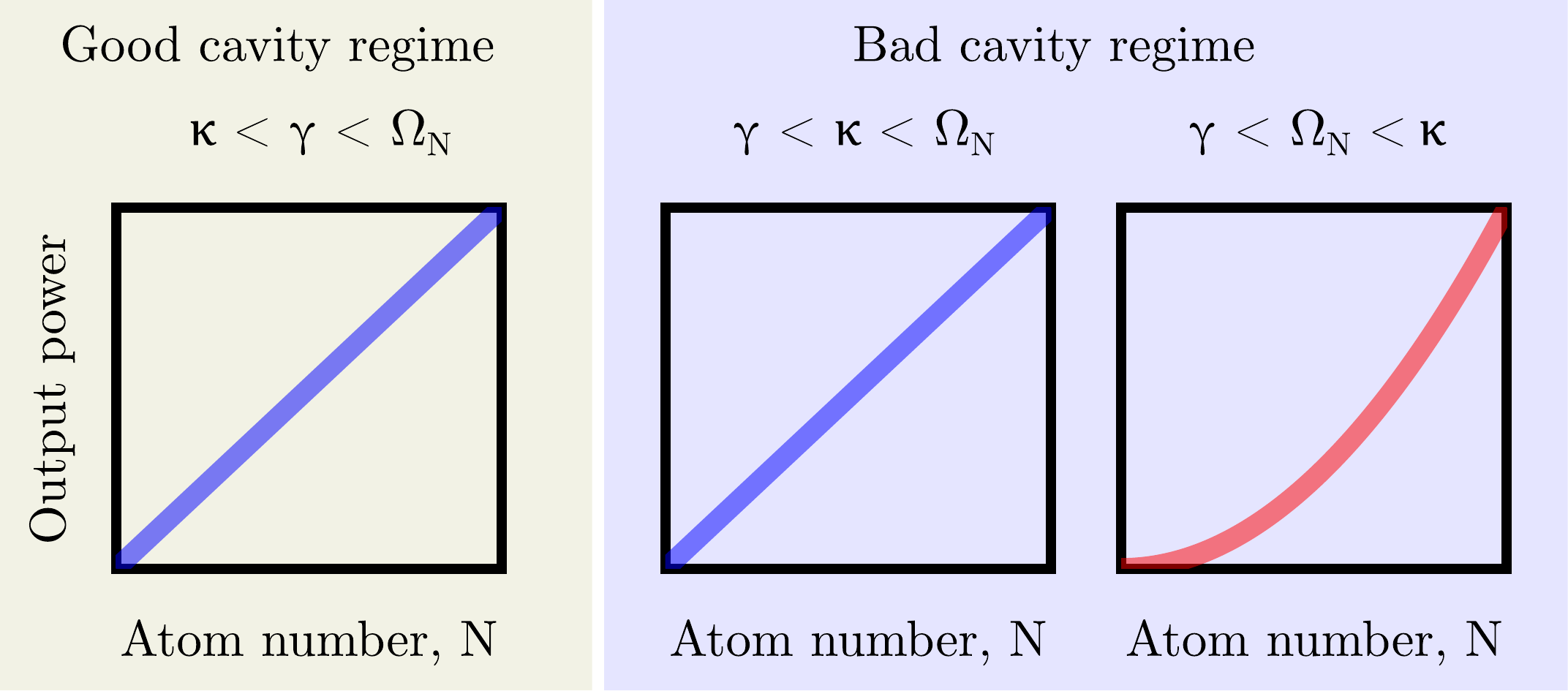}
		\caption{Three distinct lasing regimes. In our system we operate on the intersection of all three and seem to realize both $N$ and $N^2$ atom number scaling behaviors. \label{fig:regimes}}
	\end{figure}

	Our simulations indicate that the lasing occurs in three characteristic regimes determined by the ratios of the atomic decay rate $\gamma$, the cavity decay rate $\kappa$ and the collective atomic coupling factor $\Omega_N$, see Fig. \ref{fig:regimes}. We consider only $\Omega_N>\gamma$ since for $\Omega_N<\gamma$ the spontaneous atomic decay to the reservoir will dominate. In the bad cavity regime where $\gamma<\kappa$ a low coupling strength will allow the output power to scale quadratically with atom number, whereas high coupling strength will result in a linear power scaling and multiple Rabi oscillations in the atomic population during a single lasing pulse. Broadening effects can lead to higher effective decoherence rates $\Gamma_{dec}$ which must be considered.
	
	\section{System characterization}
	We characterize the lasing properties and pulse dynamics of the system by varying cavity-atom detuning and atom number. We compare the measurements with simulated experiments to verify the understanding in the numerical model. This will then allow us to draw out some behaviors from the model that are inaccessible experimentally.
	
	\subsection{Lasing Threshold}
	
	\begin{figure}[t]
		\includegraphics[width=\columnwidth]{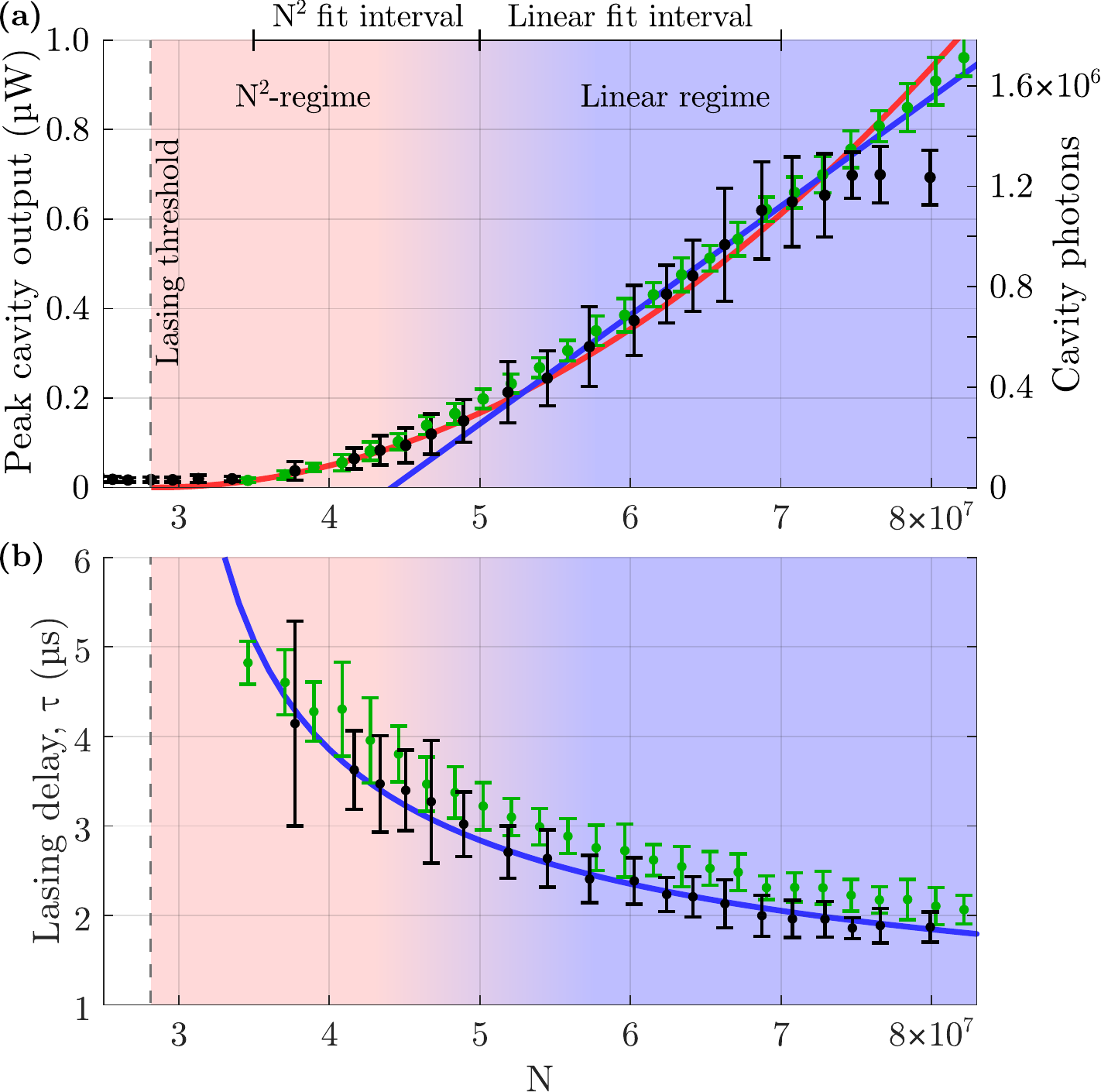}
		\caption{Primary pulse behavior. \textbf{(a)} Atom number dependency of peak cavity output power at a single mirror. Black points indicate data, whereas green points indicate simulation results. Threshold ($N_{th}$) is determined by fitting the quadratic curve (red), and the linear curve (blue) is fitted only to $N=5\textrm{ to }7\times10^7$. \textbf{(b)} Delays between pumping and lasing pulses. The delay is defined as the time interval between the end of the pumping pulse, and the peak intensity of the subsequently emitted laser pulse. The blue curve is fit using $a/\sqrt{N-N_{th}}$.\label{fig:threshold}}
	\end{figure}
	
	By varying the total number of atoms in the trap, we find the lasing threshold of the system. This gives an atom number dependency of the pulse power and the associated delay, see Fig. \ref{fig:threshold} (a). We plot the peak power of the emitted lasing pulse as a function of the atom number $N$ in the full atomic cloud for $795$ experimental runs. Each point is about $40$ binned experimental runs, with the standard deviation indicated. For low atom numbers (white region), $N\leq3\times10^7$, excited atoms decay with their natural lifetime, $\tau=22$~$\upmu$s. In an intermediate regime of $3\times10^7<N<5\times10^7$  (red region) the peak power appears to scale quadratically with the atom number, as the onset of lasing occurs. This is what we would expect according to the parameter regimes illustrated in Fig. \ref{fig:regimes}. Finally, for higher atom numbers (blue region), $N>5\times10^7$, the peak power becomes linearly dependent on the atom number, as the collective coupling $\Omega_N$ becomes much larger than $\kappa$. We show a red and blue curve fitted to their corresponding regions with an $N^2$- and $N$-scaling respectively. The green points show the results from simulation, and appear to agree well with experiment. The curves are fitted to the raw data, within the respective regions before any binning, and the lasing threshold is determined from the quadratic fit. While the lasing process does not initiate for low atom numbers due to the requirement that $C_N\gamma \gg \Gamma_{dec}$, for high $N$ the cavity field builds up sufficiently that the slowest atoms become strongly driven. At high atom numbers ($N>7\times10^7$) the cavity output seems to saturate as the assumption of linear scaling of the cavity atom number ($N_{cav}$) with respect to the MOT atom number ($N$) breaks down, and these points are not included in the linear fit. The nonzero value of the data below threshold is caused by random noise.
	
	In Fig. \ref{fig:threshold} (b) we plot the delay between the end of the pumping pulse and the associated peak in laser pulse emission. The delay has high uncertainties in the red region where we expect an $1/N$ scaling \cite{Gross}. For high $N$ we expect a $1/\sqrt{N}$ trend \cite{Kumarakrishnan} (blue), and we choose to fit this to the entire dataset. When going to low atom numbers background noise becomes increasingly important until no emission peak is visible, and the effective delay goes to infinity. Once again we show the simulation with green dots. Notice that there is a clear tendency towards longer delays in the simulation. We believe this to be caused by the fact that the model does not include spontaneous emission into the cavity mode.
	
	As the number of atoms decreases, so does the collective cooperativity, and thus the ensemble coherence build-up. The time it takes for the ensemble to phase-synchronize increases, leading to the increased delay time and associated decrease in peak intensity. The total number of photons emitted during a pulse is not constant, but scales with the collective cooperativity $C_N$, just as the peak output power in Fig. \ref{fig:threshold}.
	
	\subsection{Pulse evolution}
	We map out the time evolution as a function of atom number and cavity-atom detuning respectively. To ease interpretation we define $t=0$~s to be at the time of maximal power of each dataset in figures \ref{fig:Nvar} and \ref{fig:CavDet}. The emission-delay is then given by the distance between this maximal value at time zero and the green dots indicating the end of the pumping pulse.
	
	\subsubsection{Atom number variation}
	
	\begin{figure}[t]
		\includegraphics[width=\columnwidth]{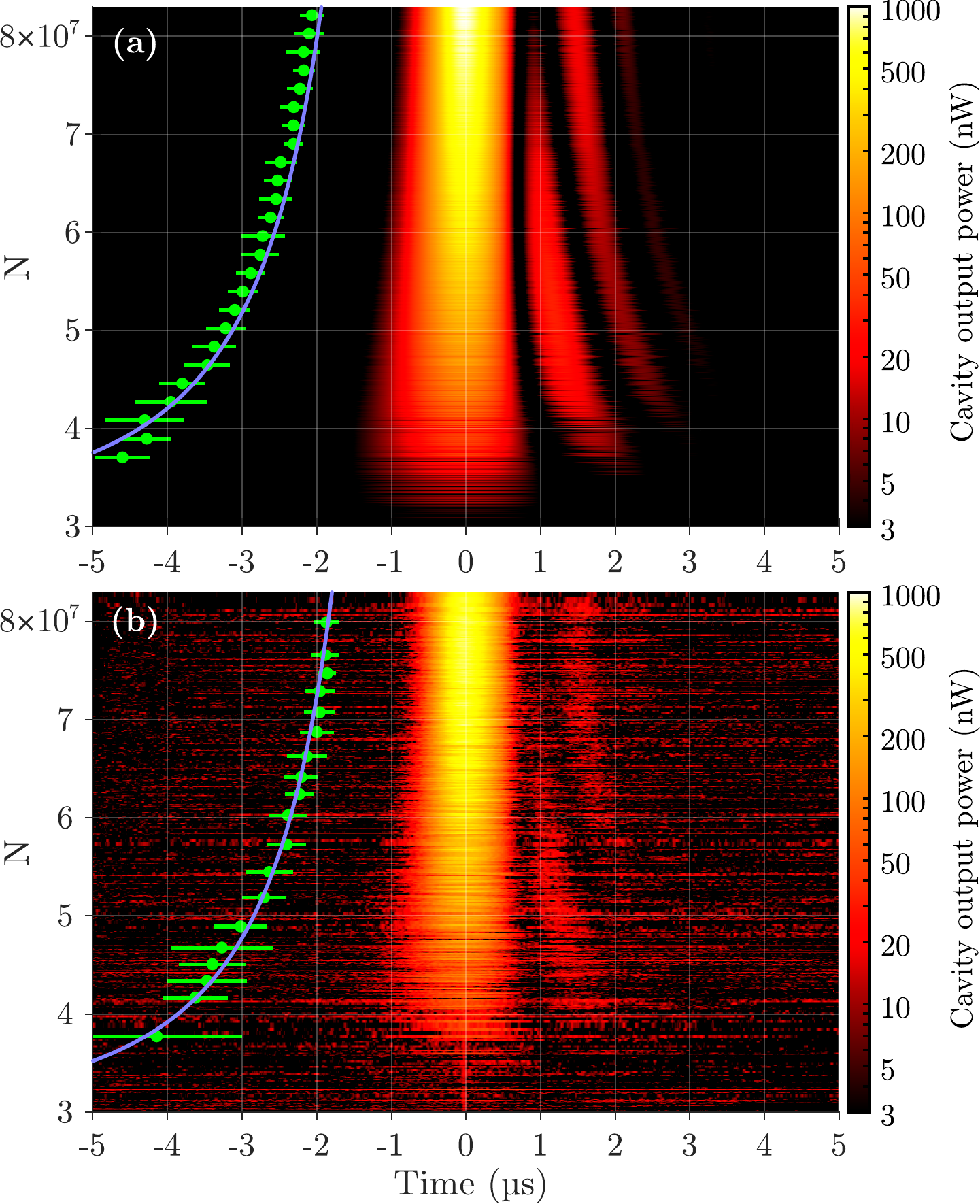}
		\caption{Atom number ($N$) dependency of the pulse dynamics.
			\textbf{(a)} Simulation results. When the cavity mode is at atom resonance, the characteristic frequency of the lasing pulse oscillatory behavior, varies weakly with the atom number. The pulse delays $\tau$ are shown with a $a/\sqrt{N-N_{th}}$-fit (blue).
			\textbf{(b)} Experimental results. The simulations are nicely replicated by experiment, with the first revival disappearing in noise at around $N=7\times10^7$. \label{fig:Nvar}}
	\end{figure}
	
	In Fig. \ref{fig:Nvar} (a) we simulate the behavior of the lasing pulse when the atom number is changed. We set the cavity-atom detuning to zero ($\Delta_{ce}=0$) and vary the atom number in the MOT while keeping the density profile constant. The green points indicate the end of the pumping pulse, and are binned in order to ease interpretation. The oscillation frequency of the output revivals decreases as the atom number is reduced, and at low atom numbers only the primary lasing pulse is visible.
	
	On Fig. \ref{fig:Nvar} (b) experimental results are shown. We vary again the atom number, allowing comparison to the simulations. A clear primary peak can be seen for atom numbers $N>3\times10^7$, and subsequent revivals of the light emission is observed. The reference light used for cavity locking shows up as a noisy background in the experimental data, and a constant offset corresponding to the mean value of the background signal has been subtracted in both figures \ref{fig:Nvar} (b) and \ref{fig:CavDet} (b). The data was recorded in sets of 50 with a randomly chosen set-point for the atom number, $N$. This means that slowly varying experimental conditions show up as slices of skewed data, but does not result in unintended biasing of the results.
	
	The oscillatory behavior is well explained by following the evolution of the atomic inversion. For high atom numbers the collective coupling in the system becomes strong enough that the photons emitted into the cavity mode are not lost before significant reabsorption takes place. This leads to an ensemble that is more than $50 \%$ excited by the end of the primary lasing pulse, and subsequent laser revival (oscillations) will thus follow. In this regime where the collective coupling rate is much larger than the cavity decay rate $\Omega_N>\kappa$ the system output is expected to behave as the central plot of Fig. \ref{fig:regimes}. At low atom numbers we get $\Omega_N<\kappa$ and light emitted by the atoms is lost from the cavity mode too fast to be reabsorbed by the atoms. As a result the primary lasing pulse creates a significant reduction in the ensemble excitation so that no further collective decay occurs, and the atoms subsequently decay only through spontaneous emission. In this regime the output power is expected to scale as $N^2$, illustrated by the rightmost graph of Fig. \ref{fig:regimes}. This is the behavior expected from an ideal superradiant system \cite{Gross, Dicke, Kumarakrishnan}.

	\subsubsection{Cavity detuning}
	
	\begin{figure}[t]
		\includegraphics[width=\columnwidth]{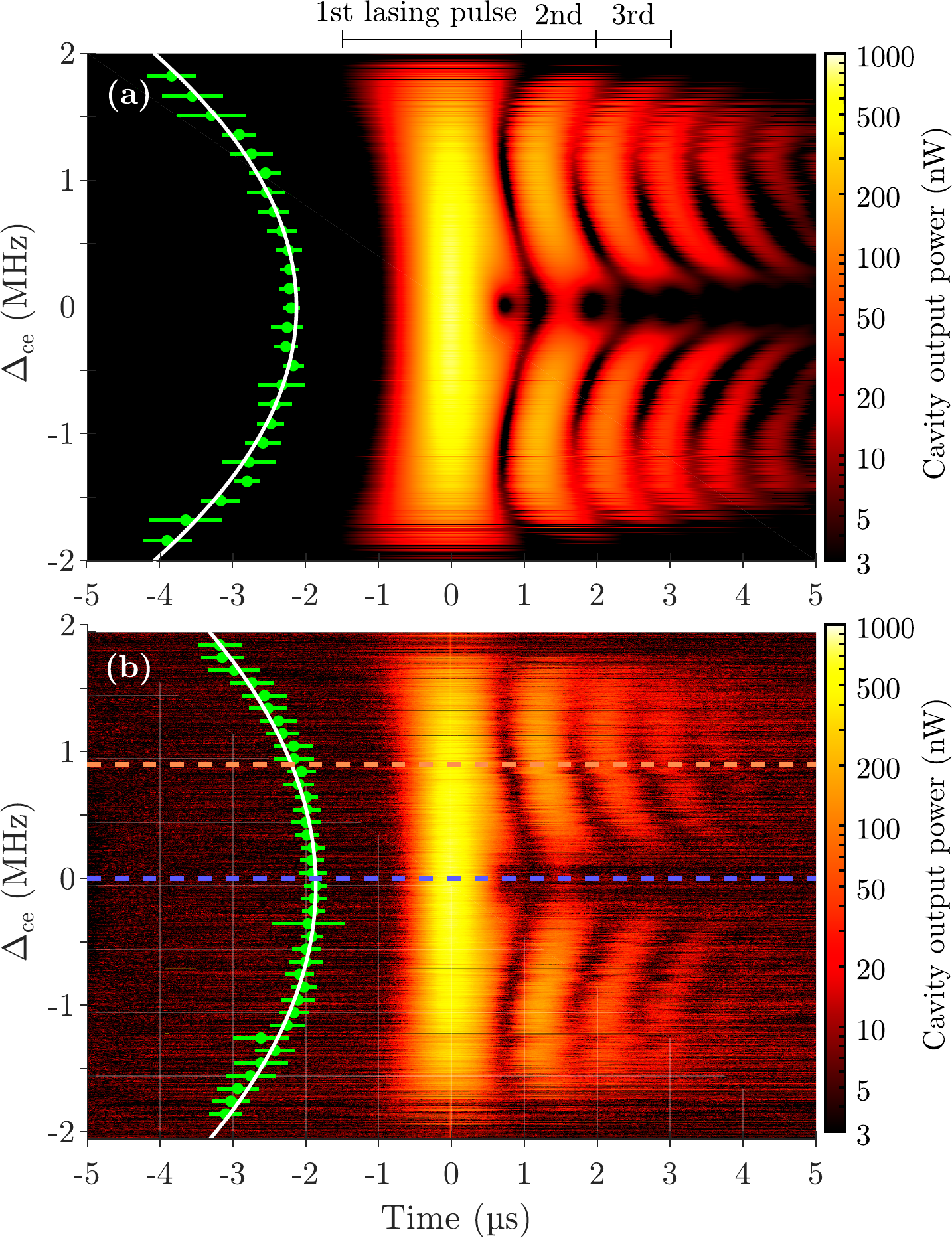}
		\caption{Cavity-atom detuning dependency of the pulse dynamics.
		The oscillatory behavior is strongly suppressed at resonance, but appears clearly once the cavity is detuned. Here at least four subsequent revivals are visible. The oscillating frequency scales with the detuning frequency and appears symmetric around zero detuning. The white line is a $\Delta_{ce}^2$-fit to the pulse delays $\tau$. The atom number is $N=7.5 \times10^7$. 
			\textbf{(a)} Simulation results. The temperature was set to $T=5$~mK.
			\textbf{(b)} Experimental results. The dashed blue and orange lines indicate the cuts shown in Fig. \ref{fig:Pulse} (a). \label{fig:CavDet}}
	\end{figure}
	
	With a constant atom number of $N=7.5\times10^7$, we now vary the cavity-atom detuning $\Delta_{ce}$ in Fig. \ref{fig:CavDet}. Here a broad range of cavity-atom detunings, up to about $\Delta_{ce}=\pm 2$~MHz is seen to facilitate lasing. At zero detuning the primary lasing feature is maximal and subsequent revivals are suppressed. The lasing pulse delay is seen to scale as $\Delta_{ce}^2$, and is thus linearly insensitive to fluctuations close to $\Delta_{ce}=0$.
	
	For non-zero detuning the oscillatory behavior of the lasing intensity is seen to increase in frequency, scaling with the generalized Rabi frequency of the coupled atom-cavity system $\Omega'_N=\sqrt{4Ng_\textit{\scriptsize{eff}}^2+\Delta_{ce}^2}$ \cite{Bochmann,Brecha}. A noticeable effect is the apparent suppression of emission revivals for any detunings $\Delta_{ce}\leq 200$~kHz.
	
	The cavity-atom detuning range for which a significant lasing pulse is produced can be broad compared to the natural transition linewidth $\gamma$. The pulse can be initiated by only a few photons in the cavity field. The inhomogeneous broadening of the atomic linewidth caused by build-up of optical power in the cavity subsequently increases the lasing range significantly. As the intensity in the cavity mode builds up, power broadening acts to increase the effective mode overlap between the field and individual atoms. This increases the effective gain in the system, both at finite and zero detuning, allowing more atoms to participate and more energy to be extracted than would have otherwise been the case. Here, the lasing range agrees well with the Doppler broadening, but for the case of a much colder atomic ensemble ($\sim\upmu$K) the Doppler broadening of the atomic transition is no longer on the same order of magnitude. Further simulations, however, indicate that the width of the cavity-atom detuning region which supports lasing remains wide in the $\upmu$K case due to power broadening.
	
	\subsubsection{Velocity-dependent dynamics}
\begin{figure}[t]
		\includegraphics[width=\columnwidth]{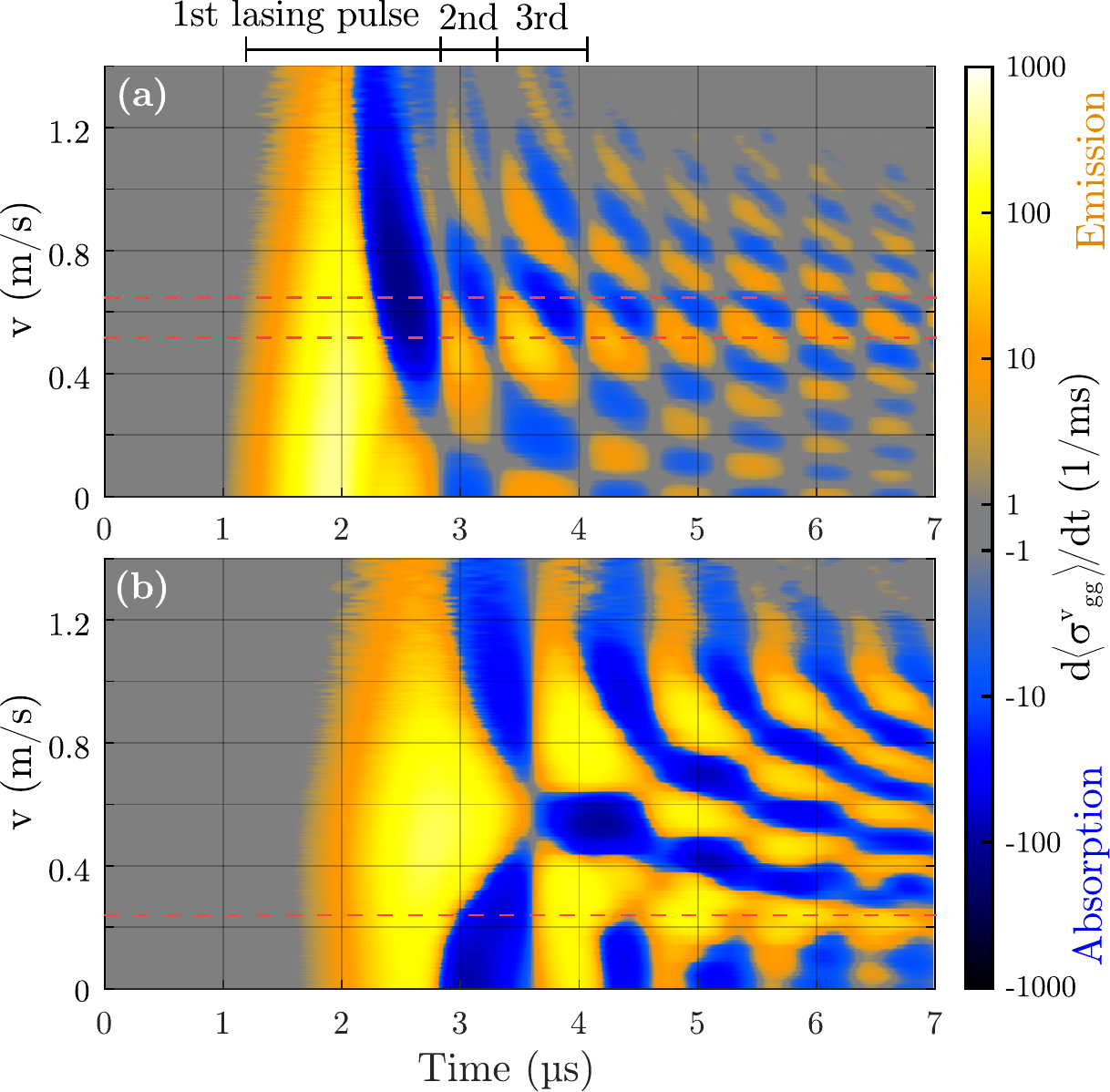}
		\caption{Velocity-dependent atomic absorption and emission during the lasing process for the case of a resonant and detuned cavity respectively. The atoms are pumped by an excitation pulse ending at time zero. Yellow colors represent emission into the cavity mode while blue represents absorption. Red dashed lines indicate velocity groups that behave consistently throughout most of the oscillations. We choose $\pm1$ as the lower cutoff of the color-scale and set all values between these to gray.
			(a) Resonant cavity. Oscillations after the primary pulse can be seen as vertical lines where $d\langle\sigma^v_{gg}\rangle/dt=0$. The atom number used here is $N=7.5\times10^7$.
			(b) The cavity field is detuned by $\Delta_{ce}=1$~MHz, and the peak emission contribution during the primary pulse comes from atoms with a speed of $v=0.6$~m/s along the cavity axis. \label{fig:velDynamics}}
	\end{figure}

	During the lasing process, the Rabi frequency of each atom will vary in time due to the changing cavity field intensity, while the atomic motion along the cavity mode leads to velocity-dependent dynamics. A typical thermal atom may move a distance of a few wavelengths during the lasing process in this temperature regime. Our simulation shows how atoms affect the lasing process differently, depending on their velocities along the cavity axis, see Fig. \ref{fig:velDynamics}. With an angle of 45$^\circ$ between the cavity axis and the pump pulse beam, the slow atoms along the cavity axis are preferentially excited during pumping. These atoms initiate the lasing process, while faster atoms may suppress the pulse amplitude by absorbing light. Different velocity classes dominate emission or absorption of the cavity photons at different times during the initial pulse and subsequent revivals. The theoretical description of the velocity-dependent behavior shown in Fig. \ref{fig:velDynamics} provides an improved qualitative understanding of the effect of having thermal atoms in the system. As atoms are cooled further, their behavior becomes increasingly homogeneous, and the asynchronous behavior of hot atoms no longer destroys the ensemble coherence.
	
	Fig. \ref{fig:velDynamics} (a) illustrates the velocity dynamics for the case of a resonant cavity, $\Delta_{ce}=0$. For a range of different velocity groups, this shows the rate of change of the atomic ground state population due to interactions with the cavity field.  Significantly more ground-state oscillations after the primary pulse are visible here than revivals in the emitted power on Fig. \ref{fig:Nvar}. Most of these oscillations see an approximately equal amount of emission and absorption, causing the energy to remain in the atomic excitations rather than being lost from decay of the cavity mode. They can, however, be seen in the phase response of the system as illustrated in Fig. \ref{fig:Pulse} (b). Eventually loss from spontaneous emission into the reservoir becomes an important decay channel. Concentrating on the slowest atoms, we see emission during the full length of the primary lasing pulse. For the subsequent pulses these atoms alternate between absorbing or emitting light. If we could isolate the light from the slowest atoms, we would thus only see every second lasing revival in the output power. For atoms with larger velocities, there will sometimes be both emission and absorption during any single pulse, and we even see the tendency of some velocity groups to consistently emit ($v=0.5$~m/s) or absorb ($v=0.65$~m/s) throughout the full process (red dashed lines on Fig. \ref{fig:velDynamics}). This indicates that even for the case of a resonant cavity the velocity groups contributing most to the emitted light during the laser revivals are not the resonant ones. In the case of a detuned cavity mode, Fig. \ref{fig:velDynamics} (b), the initial behavior is very similar. Atoms whose Doppler detuning brings them on resonance with the detuned cavity, emit throughout the primary lasing pulse, whereas others will begin to absorb. Once again some atoms ($v=0.25$~m/s) appear to emit light throughout the pulse revival oscillations. The periods of zero emission or absorption between oscillations (gray) are no longer visible, as some light is always emitted and absorbed by the atoms. The minima in the emitted power are thus no longer caused by zero emission, but rather by the cancellation between different velocity classes. This behavior corresponds well to the results of Fig. \ref{fig:CavDet} where revivals are much more pronounced in the case of large cavity-atom detuning. Future studies of the spectral properties of superradiant light in cold-atom systems, could elucidate the dependency of emitted light frequency on the finite temperature of the atoms.
	
	\section{Conclusion}
	In this paper we have investigated the behavior of an ensemble of cold atoms excited on a narrow transition and coupled to the mode of an optical resonator. The enhanced interaction provided by the cavity facilitates synchronization of the atomic dipoles, and results in the emission of a lasing pulse into the cavity mode. This realizes the fundamental operating principle for an active optical clock, where superradiant emission of laser light can be used as a narrow-linewidth and highly stable oscillator.
	
	We mapped out the emitted laser power as a function of atom number in order to identify the threshold of about $N_{cav}^{threshold}=4.5\times10^6$ atoms inside the cavity mode. Two different scalings of laser output power in the bad cavity regime are identified, and though our system is at the limit of the bad cavity regime, both regimes are realized by varying the atom number.
	
	In an attempt to quantify the decoherence effects resulting from finite atomic temperature, a Tavis-Cummings model was developed. By using detailed parameters of the pumping sequence, atomic spatial distribution and orientation, the model is seen to reproduce the experimental results to a high degree. The emitted energy from the atoms is seen to exhibit temporal Rabi oscillations as it undulates between atomic and cavity excitation. This behavior is elucidated via the numerical simulation by investigating the change in atomic excitation as a function of atomic speed throughout the lasing pulse sequence. We see that different velocity groups behave anti-symmetrically, with respect to each other. Surprisingly the velocity group mainly contributing to emission rapidly changes from resonant atoms to atoms that are more detuned with respect to the cavity. This is caused by a faster initial loss of excitation for resonant atoms. A similar effect is seen in the case of a detuned cavity. Here the atomic behavior is much more uniform across different atomic speeds, as the relation between atom number and cavity coupling becomes more homogeneous.

	This system relies on pulses of lasing from independent ensembles of atoms, which limits the pulse-to-pulse phase coherence. The phase coherence would intact between pulses if we could ensure atoms are always present in the cavity as a memory, e.g., in a continuous system. The prospect of a continuously lasing atom-cavity system based on unconfined cold atoms is intriguing because of the severe reduction in engineering requirements compared to a system based on, e.g., sequential loading of atoms into an optical lattice \cite{Kazakov}. The velocity-dependent dynamics are important in order to understand what kind of equilibrium one can expect in such a system. An investigation of the spectral characteristics of stationary atomic systems have been shown in \cite{NorciaPRX2018,Debnath}, and are promising for the transition we have used here. Investigation of the spectral properties in an unconfined ensemble will be presented in future work.

	\begin{acknowledgments}
		SAS and MT contributed equally to the scientific work. The authors would like to thank J. H. M\" uller, A. S. S\o rensen, J. Ye, J. K. Thompson, and M. Norcia for helpful discussions. This project has received funding from the European Union's Horizon 2020 research and innovation programme under grant agreement No 820404 (iqClock project), the USOQS project (17FUN03) under the EMPIR initiative, and the Q-Clocks project under the European Comission's QuantERA initiative. JWT, MRH and BTRC were additionally supported by research grant 17558 from Villum fonden.
	\end{acknowledgments}
	
	\appendix
	\section{Simulation parameters}
	
	Simulations of the system are based on numerical integration of Eqs. \ref{eq:OBE} \cite{MTthesis}. The system is initiated with all atoms in the ground state and no coherence. The atoms are randomly distributed, assuming a Gaussian density profile in each dimension, and randomly generated thermal velocities for a temperature of $T=5$~mK. These velocities are assumed constant due to negligible collision rates. The pumping is simulated by turning on the Rabi frequency $\chi_p^j$, which is calculated for each atom based on their coupling to the running-wave pump pulse and its intensity. The spatial intensity distribution is estimated based on measurements of the pump beam with a CCD camera and an optical power meter. After spatial smoothing to even out noise, the data from the CCD camera is used directly in the simulations, and correspond approximately to a slightly non-Gaussian elliptic profile with waists of $w_0^g=2.7$~mm and $w_0^m=1.5$~mm. The minor axis of the ellipse is rotated by $35^\circ$ with respect to the magnetic symmetry axis, and the power used is $P_{pump}=98.4$~mW. The simulated time evolution of the pumping pulse ignores the ramp-up and ramp-down of the AOM, and assumes a square pulse for $160$~ns. 
	
	The MOT coils impose a quantization axis for the $\Delta m = 0$ transition. The pump pulse is polarized along the MOT coil axis, and as a result, atoms near the center of the MOT field are driven less strongly by the pump pulse. In the model, this is accounted for by introducing an effective intensity driving the transition, given by $I \cdot 4y^2/(x^2 + 4y^2 + z^2)$, where the y-axis is along the MOT coil axis. In the simulations the MOT cloud center is offset by $2$~mm with respect to $y=0$ based on measurements of the energy splittings of the magnetic states. The pumping leaves the ensemble inhomogeneously excited, with the excitation being highest for atoms slightly away from the beam axis and for the slowest atoms along the beam axis. On average $85~\%$ of atoms are excited within the cavity waist at the end of the pump pulse. Atomic spontaneous decay at a rate $\gamma$ and leak of cavity photons through the mirrors at a rate $\kappa$ are accounted for by Liouvillian terms. Throughout the simulation each atom interacts with the cavity mode with different coupling rates $g_c^j$ depending on their positions relative to the Gaussian cavity mode waist ($w_0 = 0.45$~mm) and the standing wave structure. We calculate the cavity output power from one mirror (comparable to our experimental observations) by $P = \hbar \omega_c n \kappa / 2$. 
	
	The ensemble temperature is estimated from Doppler-spectroscopy, following the approach in \cite{Schaffer}, and time-of-flight measurements to be $T=5.0(5)$~mK. The atom number is estimated from fluorescence measurements and calibrated by comparing emitted photon number during a lasing pulse to the increase in fluorescence. The ensemble spatial distribution is estimated by shadow-imaging and calibrated fluorescence-distribution. The size of the atomic ensemble is measured to be $\sigma=0.9(2)$~mm assuming a Gaussian distribution, but is pulled to $\sigma=0.8$~mm in the simulations to ensure agreement between the output power levels. This is the only adjusted value, and is within the measurement uncertainties. These parameters are somewhat degenerate, but we choose to adjust $\sigma$. We attribute some discrepancy between experiment and simulation to the model assumption of a 3D Gaussian atomic distribution. The experimentally realized atomic ensemble is irregular in shape with some inhomogeneity.

\end{document}